\documentclass[11pt]{article} 
\usepackage{amsfonts,amscd,amsmath}

\def\NP#1#2{ Nucl. Phys. B #1 (#2)} 
\def\PL#1#2{ Phys. Lett. B #1 (#2)}

\def\PRL#1#2{ Phys. Rev. Lett. #1 (#2)} 
\def\PR#1#2{ Phys. Rev. D #1 (#2)}

\def\HP#1#2{ JHEP #1 (#2)} 
\def\ap{ \alpha^{\prime}} 

\def\ah{\hat\alpha}
\def\an{\alpha_0}

\newcommand{\ep}{\text e}
\newcommand{\oh}{\frac{1}{2}}

\title{A few comments on the high-energy behavior of string scattering amplitudes in warped spacetimes}
\author{Oleg Andreev\thanks{E-mail address: andreev@physik.hu-berlin.de}
\thanks{Also at Landau Institute for Theoretical Physics, Moscow, Russia}
\\ \\
Humboldt--Universit\"at zu Berlin, Institut f\"ur Physik\\
Newtonstra\ss e 15, D-12489 Berlin, Germany}

\date{}
\begin{document} 
 
\maketitle 
\begin{abstract} 
The high-energy behavior of string scattering in warped spacetimes is studied to all orders in perturbation theory. If one 
assumes that the theory is finite, the amplitudes {\it exactly} fall as powers of momentum.
\end{abstract}

\vspace{-11.5cm}
\begin{flushright}
hep-th/0402017     \\
HU Berlin-EP-04/05
\end{flushright}
\vspace{10 cm}

\section{ Introduction} 
\renewcommand{\theequation}{1.\arabic{equation}}
\setcounter{equation}{0}
Recently, the high-energy behavior of type IIB superstring amplitudes was studied in the case of warped spacetime 
geometries which are the products of $\text{AdS}_5$ with some five-manifolds [1-5]. One of the most important 
results is that of Polchinski and Strassler \cite{ps}. They proposed a scheme of evaluating high-energy fixed-angle string 
amplitudes in terms of vertex operators on a spherical worldsheet and found that the amplitudes fall as powers 
of momentum. Thus, a long standing problem on a way to the string theory description of hadronic processes was 
solved. In fact, it was already known in the seventies that the amplitudes of exclusive hadronic processes at large 
momentum transfer scale as \cite{scaling}
\begin{equation}\label{sc}
{\cal M}\sim p^{-n+4}\biggl[1+O\Bigl(\,\frac{1}{p}\,\Bigr)\biggr]
\quad,
\end{equation}
where $p$ is a large momentum scale and $n$ is a total number of hadronic constituents (valence quarks).

In this paper we extend the analysis to higher orders of string perturbation theory. We assume that perturbation theory 
in question is a topological expansion as it is in Minkowski space.

\section{ General Argument} 
\renewcommand{\theequation}{2.\arabic{equation}}
\setcounter{equation}{0}

As mentioned earlier, if spacetime geometry is chosen as the product of $\text{AdS}_5$ with a 
five-manifold K, the high-energy behavior of tree string amplitudes is hard (power law). Let us write the 
metric as 
\begin{equation}\label{metric}
ds^2=\ep^{2\varphi}\eta_{\mu\nu}dx^\mu dx^\nu+R^2d\varphi^2+R^2d\Omega_{\text K}^2
\quad,
\end{equation}
where $R$ is radius of $\text{AdS}_5$, $\eta_{\mu\nu}$ is a four-dimensional Minkowski metric. We 
assume that K doesn't provide any dimensionfull parameter except $R$. Moreover, $d\Omega_K^2$ is 
independent of $R$.

Unfortunately, full control of type IIB  string theory on curved backgrounds like $\text{AdS}_5$ is beyond 
our grasp at present. However, we will argue that the scaling behavior can be understood from a non-linear 
sigma model perspective which bypasses the known difficulty with RR backgrounds. So, the part of the 
worldsheet action which is most appropriate for our purposes is simply \footnote{We use the superspace 
notations of \cite{fms}.}
\begin{equation}\label{w-action}
S_0=\frac{1}{4\pi\ap}\int_{\Sigma_g}d^2z d^2\theta\,\,\ep^{2\varphi}\eta_{\mu\nu}
\bar DX^\mu DX^\nu
\quad,
\end{equation}
where $\Sigma_g$ is a closed Riemann surface of genus $g$. The simplest vertex operators predicated on 
this form of action are $\epsilon_{\mu\nu}(p)\bar DX^\mu DX^\nu \ep^{ip\cdot X}$ dressed 
by some $\ep^{-\Delta\varphi}{\cal O}_\Delta(\Omega_{\text K})$.\footnote{To be precise, this form 
only includes the dominant term in the limit of large $\varphi$. Subdominant terms result in 
$\frac{1}{p}$-corrections in Eq.\eqref{sc}, so we suppress them. For more discussion of vertex operators, 
see, e.g., \cite{ts} and references therein.} Then, integrated over the worldsheet $\Sigma_g$,
\begin{equation}\label{V}
V_{\Delta,p}= \int_{\Sigma_g}d^2z d^2\theta\,\,
\epsilon_{\mu\nu}(p)\bar DX^\mu DX^\nu \ep^{ip\cdot X}\ep^{-\Delta\varphi}
{\cal O}_\Delta(\Omega_{\text K})
\quad.
\end{equation}
For reasons that will soon become apparent, we restrict values of $\Delta$'s to positive integers. We also 
discard quantum numbers which are due to the manifold K.

String scattering amplitudes are defined as expectation values of the vertex operators. In the problem of interest, 
a reasonable first guess for a $g$-loop amplitude of the $2\rightarrow 2$ scattering process is \footnote{We omit the 
explicit dependence on the string coupling constant $g_s$. For a discussion of this issue,  see \cite{ps,a}.}
\begin{equation}\label{amp}
\delta^{(4)}(p_A+\dots +p_D){\cal M}_g(AB\rightarrow CD)=
\langle\prod_{i=A,\dots ,D} V_{\Delta_i,p_i}\,\rangle
\quad,
\end{equation}
where the bracket means integrals over the matter fields, ghost fields as well as moduli space of closed 
Riemann surfaces of genus $g$. We don't know the precise details about these integrals. However, 
what is only important for our purposes is an integral over the bosonic zero modes of $X$'s and $\varphi$. 
Its explicit form is given by \footnote{For a discussion of path integral measure within the sigma model approach, 
see, e.g., \cite{amt}.}
\begin{equation}\label{zero}
\int d^4x\int^{+\infty}_{-\infty}d\varphi\,\ep^{4\varphi}
\quad.
\end{equation}
After performing the integration, the amplitude takes the form
\begin{equation}\label{amp1}
{\cal M}_g(AB\rightarrow CD)=\oh\Gamma (2-\Delta/2)
\langle \,\Bigl[\frac{1}{4\pi\ap}\int_{\Sigma_g}d^2z d^2\theta\,\,
\ep^{2\varphi}\eta_{\mu\nu}\bar DX^\mu DX^\nu\Bigr]^{\frac{\Delta}{2}-2}
\prod_{i=A,\dots ,D}V_{\Delta_i,p_i}\,\rangle^\prime
\quad,
\end{equation}
where $\Delta=\Delta_A+\dots +\Delta_D$. The prime means that the zero modes were integrated out.  

One thing about \eqref{amp1} may be disturbing. It seems that the amplitude is divergent for even $\Delta$'s because  
the $\Gamma$-function prefactor develops poles.\footnote{In fact, the integral over $\varphi$ diverges at 
$\varphi=-\infty$. This divergence is due to the net factor $\ep^{-\Delta\varphi}$ coming from the vertex operators.} 
The answer to this puzzle is simple: the integration over the matter fields cancels the factor out. We will discuss this later.

We are interested in the hard scattering limit. Its kinematics is very special: there is one independent parameter - the 
Mandelstam variable $s$, while the others are functions of it and fixed scattering angle $\phi$. This allows us to 
easily determine the dependence on $s$ of the amplitude for the scalars (dilatons) whose masses are much less than $s$. 
In this case we get from \eqref{amp1} by rescaling $X=\tilde X/\sqrt{s}$, $p_i=\sqrt{s}\,\tilde p_i$
\begin{equation}\label{dil}
{\cal M}_g^{(dil)}=c_g^{(dil)}\biggl(\frac{1}{\sqrt{s}}\biggr)^{\Delta-4}
\quad.
\end{equation}
Here $c_g^{(dil)}$ is given by the right hand side of Eq.\eqref{amp1} with the $X$'s and $p_i$'s replaced by the 
$\tilde X$'s and $\tilde p_i$'s. Its independence on $s$ can be understood from general reasoning. $c_g^{(dil)}$ is a 
function of the Mandelstam variables, defined in terms of rescaled momenta, which are independent of $s$. 

In the case of hadrons, taken as bound states of spinless constituents, this would be the end of the story if the parameters 
$\Delta_i$ were related to the numbers of constituents in the corresponding hadrons as $\Delta_i=n_i$ \cite{ps}. 
Clearly the existence of spin-2 states (gravitons) as it follows from the form of the vertex operators \eqref{V} provides a strong 
objection to spinless constituents only. Because of spinning constituents, each vertex operator contributes an additional 
factor $(\sqrt{s})^{{\text S}_i}$, where ${\text S}_i$ means its four-dimensional spin. For the graviton vertex operator, 
it gives $(\sqrt{s})^2$. \footnote{As in QCD \cite{bf}, one can 
think of $\epsilon_{\mu\nu}^{(grav)}$ as $\bigl(\prod_i u_i\bigr)_{\mu\nu}\epsilon (p)$, where $\epsilon (p)$ 
plays the same role as the dilaton wave function and $u_i$ are wave functions of free spinning constituents. Note that in QCD 
the $u_i$'s are the free spinors normalized as $\sum_{\text{spin}}u\bar u=\gamma\cdot p+m$. } As a result, we get 
 \begin{equation}\label{grav}
{\cal M}_g^{(grav)}=c_g^{(grav)}\biggl(\frac{1}{\sqrt{s}}\biggr)^{\Delta+4}
\quad.
\end{equation}
Finally, the desired result is obtained by identifying the $(\Delta_i-2)$'s with the numbers of constituents in the 
corresponding hadrons as in \cite{ps}.

Having established the scaling behavior of the $g$-loop amplitude, we turn to a perturbation series. Assuming that 
perturbation theory in question is a topological expansion  and collecting together the results for each order, the amplitude 
is simply
\begin{equation}\label{full-g}
{\cal M}(AB\rightarrow CD)=\biggl(\frac{1}{\sqrt{s}}\biggr)^{n-4}\,\sum_{g=0}^\infty c_g(g_s, \phi )
\quad,
\end{equation}
where $n=n_A+\dots+n_D$. Unfortunately, we can not, with our present methods, determine the explicit form of the 
coefficients $c_g(g_s, \phi )$. 
 
At this point, a couple of short remarks is in order:
\newline(i) Our analysis can be easily generalized to include surfaces with boundaries. In this case two new features are 
especially noteworthy. First, there are vertex operators associated with boundaries. The simplest one to be suggested is 
a vector state $\epsilon_\mu(p)DX^\mu\ep^{ip\cdot X}$ dressed by 
$\ep^{-\Delta\varphi}{\cal O}_\Delta(\Omega_{\text K})$. Second, perturbation theory is determined by the 
Euler number $\chi$. Thus, the amplitude takes the form
\begin{equation}\label{full-gm}
{\cal M}(AB\rightarrow CD)=\biggl(\frac{1}{\sqrt{s}}\biggr)^{n-4}\,\sum_{\chi=2}^{-\infty} c_\chi(g_s, \phi )
\quad.
\end{equation}
\newline(ii) Certainly, our derivation of the scaling behavior is valid for any value of the radius of $\text{AdS}_5$ 
or, equivalently, for any value of the 't Hooft coupling.

\section{Simplified Model} 
\renewcommand{\theequation}{3.\arabic{equation}}
\setcounter{equation}{0}

Other approaches to the problem result in the scaling behavior at the tree level \cite{ps,a}. A notable difference is the absence 
in those of non-zero modes of the $\varphi$ and $\Omega$ fields. Nevertheless, as follows from our discussion these 
non-zero modes  do not play a crucial role in the derivation of the scaling behavior and therefore may be discarded. 
So, it is quite natural to pursue this line of thought further.

A reasonable first guess for a $g$-loop amplitude within the simplified models is  
\begin{equation}\label{am}
{\cal M}_g(AB\rightarrow CD)=\int d^5\Omega d\varphi\,\ep^{4\varphi}\, 
{\cal A}_g (AB\rightarrow CD) \prod_{i=A,\dots ,D} \ep^{-\Delta_i\varphi}\,\psi_i(\Omega)
\quad,
\end{equation}
where ${\cal A}_g(AB\rightarrow CD)$ is the standard string $g$-loop amplitude with $\ap$ replaced by 
$\ah=\ap \ep^{-2\varphi}$ and $\psi_i$ are normalized eigenfunctions of the Laplace operator on K. 

For the sake of simplicity, let us specialize to the case of scalars, e.g., the dilatons. On the one hand side, in the hard scattering 
limit the scaling behavior of the amplitude can be derived by rescaling $\varphi$ as 
$\varphi\rightarrow\varphi+\oh\ln\ap s$. At the tree level this method was used in \cite{a} but it also works 
for \eqref{am}. 

On the other hand, it is known \cite{gross1} that ${\cal A}_g$ is well approximated by its saddle point expression 
${\cal A}_g\approx (\ah s)^a\ep^{-b\ah s}$, where $a,b$ are some positive functions of $g$ and $\varphi$. This makes 
it possible to perform the integral over $\varphi$ explicitly. As expected, the amplitude falls as a power of $\sqrt s$. A 
by-product of the integration is $\Gamma(a-2+\Delta /2)$. Unlike the $\Gamma$-function prefactor 
in \eqref{amp1}, it is now finite. The point is that the integrand becomes non-singular at $\varphi=-\infty$ after the 
integration over the matter fields and moduli generates a factor $\ep^{-bs\ap\ep^{-2\varphi}}$ which 
dumps $\ep^{-\Delta\varphi}$.

It is worth noting that in Minkowski space the saddle point evaluation is valid for a given order of perturbation theory 
in the limit of large $\ap s$ \cite{gross1}. Thus, a small parameter in question is $1/\ap s$. This is not the case for 
warped geometries. Let us first try to get a heuristic understanding of what happens. ${\cal A}_g$ is now expanded in 
powers of $1/\ah s$. The rest integral is dominated by $\varphi\sim\varphi_\ast$, where $\ap s\,\ep^{-2\varphi_\ast}\sim
\Delta$. Thus, we end up with an expansion in $1/\Delta$. This means that $\Delta$ should be large in order for 
the saddle point evaluation to be valid. Actually one can come to the same conclusion on general grounds. The 
dependence in the dilaton amplitude of $\ap s$ is absorbed into the redefinition of $\varphi$. Then, as it follows from 
our ansatz \eqref{am}, the amplitude becomes a function of $\Delta$ only. If a saddle exists, it may be a good 
approximation only in the limit of large $\Delta$. The point is that the $\Delta_i$'s are restricted to positive integers, 
so $\Delta$ is an integer bounded from below, and the limit of small $\Delta$ doesn't exist.\footnote{Strictly speaking, 
it assumes a single hard process. Landshoff diagrams may result in non-integer $\Delta$'s. }

A final remark: it was shown in \cite{gkp} that states with large quantum numbers are described by special classical 
solutions of the $\text{AdS}_5\times \text{S}^5$ non-linear sigma model in the limit of large 't Hooft coupling. Certainly, no 
derivation of the high-energy fixed-angle scattering amplitudes of states with large $\Delta_i$'s from classical 
solutions is known. But we believe that the preceding comments are significant hints that it can be done, and this issue is 
worthy of future study. 

\section{Concluding Comments } 
\renewcommand{\theequation}{4.\arabic{equation}}
\setcounter{equation}{0}

(i) Here we are considering type IIB string theory, however, similar results will hold for the other string theories. 
In fact, warped geometry in spacetime is sufficient to formally ensure that scattering amplitudes are hard in the 
high-energy limit at fixed angle. For instance, in the case of bosonic string one can get the scaling behavior of 
amplitudes by repeating the arguments of section 2 for the bosonic part of the worldsheet action. 

Actually, there might be gaps in the above reasoning. First, one must show that the corresponding background is conformal. 
From this point of view type IIB is of course preferable to the others. Second, the coefficients $c_g,\,c_\chi$ must be finite or, 
in other words, the theory must be perturbatively renormalizable. Third, the series \eqref{full-g}, \eqref{full-gm} must 
converge. Even if we expect that superstring theory is finite at a given order of perturbation theory, there is no guarantee for 
convergence. For instance, a rapid growth of the volume of moduli space might be the reason for divergence \cite{gp}. 
As noted earlier, we can not, with our present methods, determine the explicit form of the coefficients $c_g,\,c_\chi$ and, 
therefore, address the issue of convergence. 

Thus, our general statement is that if the theory is finite, the amplitudes {\it exactly} fall as powers of momentum.
\newline (ii) A strong belief is that the geometry given by \eqref{metric} is valid in the limit of large $\varphi$ or, equivalently, 
large $r$, where $r=R\ep^\varphi$. At smaller values of $r$ it is somehow deformed. So, it is of some interest to evaluate 
corrections to the scaling. To do so, let us first consider the simplified model of section 3. In order for the saddle point 
evaluation to be consistent with the deformation of geometry, we should require that $\varphi_\ast$ is large. This means 
that $\ap s\gg\Delta$. Truncating then the geometry at some small $r=r_0$, we estimate the correction to ${\cal M}_g$ as 
\begin{equation}\label{corr}
\int_0^{r_0}dr\,r^{3-\Delta}{\cal A}_g(AB\rightarrow CD)\sim(\an s)^a\ep^{-b\an s}
\end{equation}
and we note that it is just the soft string amplitude with $\ap$ replaced by $\an=\ap R^2/r^2_0$. It is indeed subleading to the 
hard amplitude. 

Returning to the settings of section 2, one thing that can help with understanding of what happens in the hard scattering limit is 
some analogue between $\varphi$ and the Liouville field of 2d gravity. It was suggested by Polyakov \cite{pol} and exploited in 
\cite{a} for deriving the scaling behavior of the amplitudes. Let us pursue this point of view further. Assuming that $\varphi$ is 
slowly varying, we can consider the notion of a low-energy effective action. This action occurs from the right hand side of 
Eq.\eqref{amp} after the integration over the matter fields, ghosts and moduli. It is easy to find an effective potential along the 
lines of section 3. It is given by \footnote{We drop possible linear terms. From the viewpoint of 2d gravity these correspond 
to the dilaton background.}
\begin{equation}\label{pot}
V_{eff}(\varphi)=\mu\,\ep^{-2\varphi}
\quad,
\end{equation} 
where $\mu$ is an effective cosmological constant linearly depending on $s$. Thus, the effective potential 
suppresses the path integral for large negative $\varphi$. The effect becomes stronger with the growth 
of $s$. Finally, only large positive values of $\varphi$ will be allowed. The known 2d gravity analogue of this is 
the ``Liouville wall'' which keeps the theory in the weak coupling regime.

\vspace{.25cm} {\bf Acknowledgments}

\vspace{.25cm} 
We would like to thank S. Brodsky and G.F. de Teramond for useful discussions concerning this subject, and H. Dorn 
and A.A. Tseytlin for comments on the manuscript. The work is supported in part by DFG under Grant No. DO 447/3-1 
and the European Commission RTN Programme HPRN-CT-2000-00131.



\end{document}